\documentclass[showpacs,preprintnumbers,english,amsmath,amssymb]{revtex4}
\usepackage[T1]{fontenc}
\usepackage[latin1]{inputenc}
\usepackage{graphicx}
\usepackage{amssymb}
\usepackage{verbatim}
\usepackage{epic,eepic}
\usepackage{babel}
\usepackage{color}

\begin{document}

\title{Coulomb corrections to density and temperature of bosons in heavy ion collisions}

\author{Hua Zheng$^{a,b)}$\footnote{Electronic address: zhengh@tamu.edu}, Gianluca Giuliani$^{a)}$ and Aldo Bonasera$^{a,c)}$}
\affiliation{
a)Cyclotron Institute, Texas A\&M University, College Station, TX 77843, USA;\\
b)Physics Department, Texas A\&M University, College Station, TX 77843, USA;\\
c)Laboratori Nazionali del Sud, INFN, via Santa Sofia, 62, 95123 Catania, Italy.}




\begin{abstract}
A recently proposed method, based on quadrupole and multiplicity fluctuations in heavy ion collisions, is modified in order to take into account distortions due to the Coulomb field. This is particularly interesting for bosons such as $d$ and $\alpha$ particles, produced in heavy ion collisions. We derive the temperatures and densities seen by the bosons and compare them to results of similar calculations for fermions. The resulting energy densities agree rather well with each other and with the one derived from neutron observables. This suggests that a common phenomenon, such as the sudden opening of many reaction channels and/or a liquid gas phase transition, is responsible for the agreement.
\end{abstract}


\maketitle

\section{Introduction}
Important information   about the Nuclear Equation of State (NEOS) can be obtained by colliding heavy ions \cite{toro1, bao1, lynch1,aldo1}. The task is not easy since we have to deal with a microscopic dynamical system. Non-equilibrium effects might be dominant and we have to derive quantities, such as density, temperature and pressure to constrain the NEOS. To minimize for non-equilibrium effects, we define different physical quantities in the transverse direction, a practice commonly  used in relativistic heavy ion collisions and other fields. Recently we have proposed a method to determine density and temperature from fluctuations \cite{sara, hua1, hua2, hua3, hua4}. The reason for looking at fluctuations, especially in the perpendicular direction  to the beam axis, is because they are directly connected to temperature through the fluctuation-dissipation theorem \cite{landau}, for instance. Of course, the system might be chaotic but not ergodic, in which case  fluctuations should give the closest possible approximation to  the 'temperature' reached during the collisions. Quadrupole fluctuations (QF) \cite{sara} can be easily linked to the temperature in the classical limit. If the system is classical and ergodic, the temperature determined from QF and from the slope of the kinetic energy distribution function \cite{khuang}, i.e. the number of ions in the energy interval $dE$, of the particles should be the same. In the ergodic case, the temperature determined from isotopic double ratios \cite{albergo} should also give the same result.  This is, however, not always observed, which implies that the system is neither  ergodic, nor classical. Since we are dealing with bosons  and/or fermions it can be necessary to go beyond classical approximation \cite{hua1, hua2, hua3, hua4}. In a previous work \cite{hua4} we have discussed Coulomb corrections to fermion fluctuation temperatures. In this paper we will concentrate on bosons, i.e. $\alpha$ and $d$ particles.  The boson case was also discussed in \cite{hua3} but without Coulomb corrections. It is well known that ideal Bose gases give unphysical results near and below the critical point. These problems are mitigated or completely resolved when the bosons experience some repulsive potential.  This is surely the case for $\alpha$ and $d$ where, at least, the Coulomb repulsion must be included. For bosons it is not possible to disentangle the 'temperature' from the critical temperature $T_c$, and  thus the density \cite{hua3}. Since we have two unknowns, we need another observable, which depends on the same physical quantities. In \cite{hua1, hua2, hua3} we have proposed to look at multiplicity fluctuations (MF) which also  depends on $T$ and $\rho$ of the system in a way typical for fermions \cite{hua1, hua2} or bosons \cite{hua3}. The application of these ideas in experiments has produced interesting results such as a sensitivity of the temperature to the symmetry energy \cite{alan1}, fermion quenching \cite{brian} and the critical $T$ and $\rho$ in asymmetric matter \cite{justin}. Very surprisingly, the method based on quantum fluctuations \cite{justin} gives values of $T$ and $\rho$ very similar to those obtained using the double ratio method and coalescence \cite{qin, kris, ropke} and gives a good determination of the critical exponent $\beta$. This raises the question why, in some cases, different methods give different values \cite{tsang1, tsang2}, while in other cases the same values are obtained.

 In \cite{sara} the classical temperature derived from QF gave different values for different isotopes. In \cite{hua4} we showed that Coulomb corrections result in similar $T$ for different nuclei having the same mass number. We also showed that the Coulomb repulsion of different charged particles can distort the value of the temperature obtained from QF, which depends on kinetic properties.  On the other hand, MF for different particles seem to be unaffected by Coulomb effects as we have discussed in \cite{hua3, hua4}. Also the obtained values, say of the critical temperature and density, might also be influenced by Coulomb effects ( as well as by finite size effects). For these reasons, it is desirable to correct for these effects as best as possible.  It is the goal of this paper to propose a method to correct for Coulomb effects in the exit channel of produced charged particles. In order to support our findings, we will compare our results to the neutron observables, which should not depend directly on the Coulomb force. Neutron distributions and fluctuations are not easily determined experimentally. Thus we will base our considerations on theoretical simulations using the Constrained Molecular Dynamics (CoMD) approach \cite{comd}. These simulations have already been discussed in \cite{hua1, hua2, hua3, hua4} for ${}^{40}Ca+{}^{40}Ca$ at $b=1fm$ and for beam energies ranging from $4$ MeV/nucleon to $100$ MeV/nucleon in the laboratory system. About 250,000 events for each case have been generated.
\begin{figure}
\centering
\includegraphics[width=0.5\columnwidth]{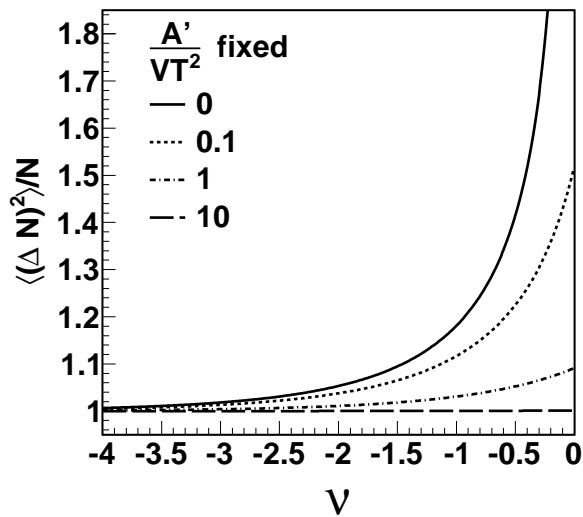}
\caption{The multiplicity fluctuations versus $\nu$ with fixed $\frac{A'}{VT^2}$ in Eq. (\ref{quancoulcorrmf}). Different lines refer to different values of $\frac{A'}{VT^2}$.}
\label{try1}
\end{figure} 

Let us imagine that we have a charged particle, say a deuteron with charge $Z_d$, leaving a system of charge $Z_s$, mass $A$ in a volume $V$. The particle momentum is ${\bf p_i}$, and it gets accelerated by the Coulomb field to the final momentum ${\bf p_f}$. Assuming a free particle  wave function for that particle, the Coulomb field becomes \cite{hua4, povh}:
\begin{eqnarray}
V(q) &=& \langle \psi_f|H_{int}|\psi_i\rangle \nonumber\\
&=& \frac{Z_de}{V}\int e^{-i{\bf p_f\cdot x}/\hbar} \phi({\bf x})e^{i {\bf p_i \cdot x}/\hbar} d^3x\nonumber\\
&=& \frac{1.44\times 4\pi  \hbar^2Z_dZ_s }{q^2 V}F({\bf q}), \label{formfactor}
\end{eqnarray}
where ${\bf q=p_i-p_f}$, $\phi({\bf x})$ is the Coulomb potential of the source, $F({\bf q})$ is the form factor. This is similar to that assumed in the density determination of the source in electron-nucleus scattering \cite{povh}. To make calculations feasible, we will assume that ${\bf p_i}$ is negligible, which is not a bad approximation at low energies or temperatures such that most of the charged particle acceleration is due to Coulomb effects. At high excitation energies we expect Coulomb effects to be negligible \cite{aldo2, huang} since the source is at at high temperature and relatively low density. In fact we have seen in previous calculations \cite{hua1, hua2, hua3, hua4} that charged and uncharged particles produced in the collisions at high energies give similar values of $T$ as expected. For simplicity we will also assume that the form factor is equal to $1$. More involved numerical calculations and higher statistics  are needed in order to fully explore the form factor $F({\bf q})$. Unfortunately, those calculations are presently beyond our numerical capabilities.

\begin{figure}
\centering
\includegraphics[width=0.5\columnwidth]{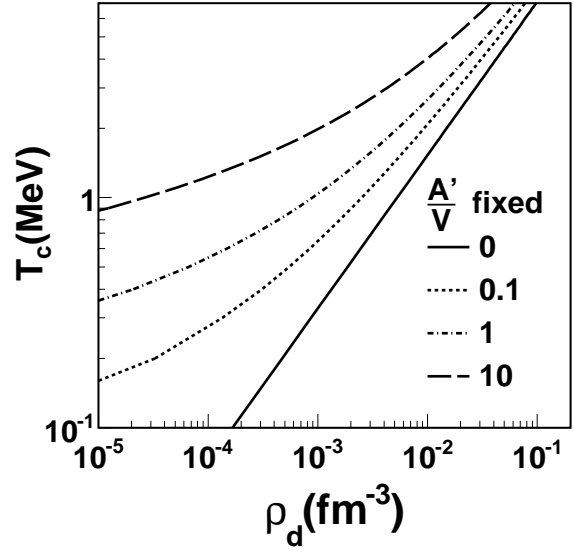} 
\caption{Critical temperature versus density with fixed $\frac{A'}{V}$. We take $d$ as an example.}
\label{try2}
\end{figure}

The reason for essentially making a Fourier transform of the Coulomb field, is because the distribution function is modified by the factor \cite{landau}:
\begin{equation}
f(p)\propto \exp[-\frac{R_{min}}{T}]\propto \exp[-\frac{V(q=p)}{T}].
\end{equation}
Using this result, we can estimate modifications to physical quantities in the classical and quantum cases. The classical case is interesting because it gives smaller temperatures for different fragments, very close to the neutron values, as was discussed in \cite{hua4}.  The correction to the distribution function  is, as we will show below, crucial for bosons. 

\section{Quantum case}
In this work we will restrict the results to the $d$ and $\alpha$ cases. In the quantum case, considering the Coulomb correction, the QF can be obtained from:
\begin{equation}
\langle \sigma_{xy}^2 \rangle=(2mT)^2\frac{4}{15}\frac{\int_0^\infty dy y^\frac{5}{2} \frac{1}{e^{y+\frac{A'}{yVT^2}-\nu}-1}}{\int_0^\infty dy y^\frac{1}{2} \frac{1}{e^{y+\frac{A'}{yVT^2}-\nu}-1}},
\label{quadcoulcorrqf}
\end{equation}
where $A'=\frac{1.44\times4\pi\hbar^2 q_1q_2}{2m}$ and $\nu=\frac{\mu}{T}$. On the same ground we can derive the MF as:
\begin{equation}
\frac{\langle(\Delta N)^2\rangle}{N}=\frac{\int_0^\infty dy y^\frac{1}{2}\frac{e^{y+\frac{A'}{yVT^2}-\nu}}{(e^{y+\frac{A'}{yVT^2}-\nu}-1)^2}}{\int_0^\infty dy y^\frac{1}{2}\frac{1}{e^{y+\frac{A'}{yVT^2}-\nu}-1}}.
\label{quancoulcorrmf}
\end{equation}
A detailed derivation of Eqs. (\ref{quadcoulcorrqf}, \ref{quancoulcorrmf}) is given in the appendix \ref{quanboson}. We introduce three variables $T, V$ and $\nu$ into Eqs. (\ref{quadcoulcorrqf}, \ref{quancoulcorrmf}). This means that to solve those equations we need one more condition. We choose the average multiplicity:
\begin{equation}
N =\frac{gV}{h^3}4\pi\frac{(2mT)^\frac{3}{2}}{2}\int_0^\infty dy y^\frac{1}{2}\frac{1}{e^{y+\frac{A'}{yVT^2}-\nu}-1}.
\label{avn}
\end{equation}

\begin{figure}
\centering
\includegraphics[width=0.5\columnwidth]{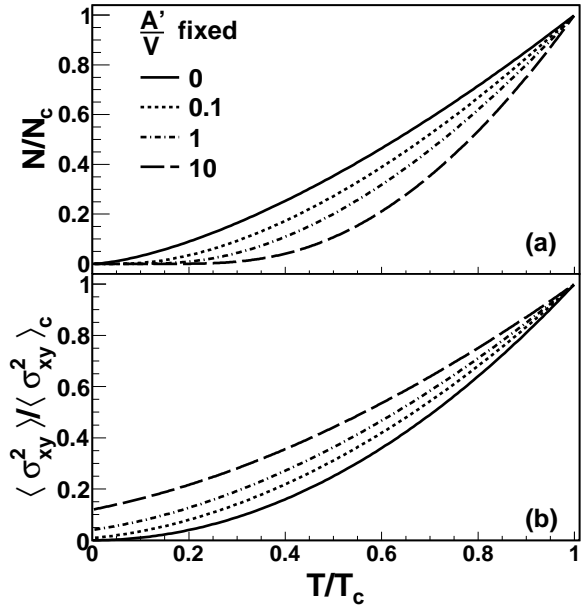}
\caption{N of bosons (a) and quadrupole fluctuations (b) divided by their respective values at the critical point vs $T/T_c$.  This result  is, to a very good approximation,  independent of the particle type (i.e. $\alpha$ or $d$) at one fixed density.}
\label{try3}
\end{figure} 
Again the detailed derivation is given in the appendix \ref{quanboson}. Those equations can be solved numerically. In figure \ref{try1} we plot the $\frac{\langle(\Delta N)^2\rangle}{N}$ vs $\nu$ with fixed $\frac{A'}{VT^2}$ in Eq. (\ref{quancoulcorrmf}). One can see that the  $\frac{\langle(\Delta N)^2\rangle}{N}$ is always larger than 1. When $\frac{A'}{VT^2}=0$, i.e. no Coulomb correction, the $\frac{\langle(\Delta N)^2\rangle}{N}$ recovers the ideal Bose gas result when $T>T_c$ \cite{hua3} and it diverges at the critical point. For $T<T_c$,  $\nu=0$ and we get a Bose condensate. An interesting question is what the energy of the condensate is in the case with Coulomb repulsion. Of course, we should first stress that we are dealing with finite systems. The Coulomb term gives a correction which has some similarities with the repulsive potential used in realistic Bose gases as first proposed by Bogoliubov \cite{landau,lee1, lee2}. For simplicity we  assume that the ground state energy of the condensate is that of a uniformly charged sphere of radius $r$. However, in the following we do not need any information on the ground state of the system and we have included this in the discussion just for completeness. We can rewrite Eq. (\ref{avn}) as
\begin{equation}
\rho = \frac{g}{h^3}4\pi\frac{(2mT)^\frac{3}{2}}{2}\int_0^\infty dy y^\frac{1}{2}\frac{1}{e^{y+\frac{A'}{yVT^2}-\nu}-1}.
\end{equation}
In figure \ref{try2}, we plot the critical temperature versus density for different values of $\frac{A'}{V}$. For a fixed density and volume, the Coulomb energy is larger and the critical temperature is higher. This probably provides a larger chance for bosons to reach the lowest energy state.
\begin{figure}
\centering
\includegraphics[width=0.5\columnwidth]{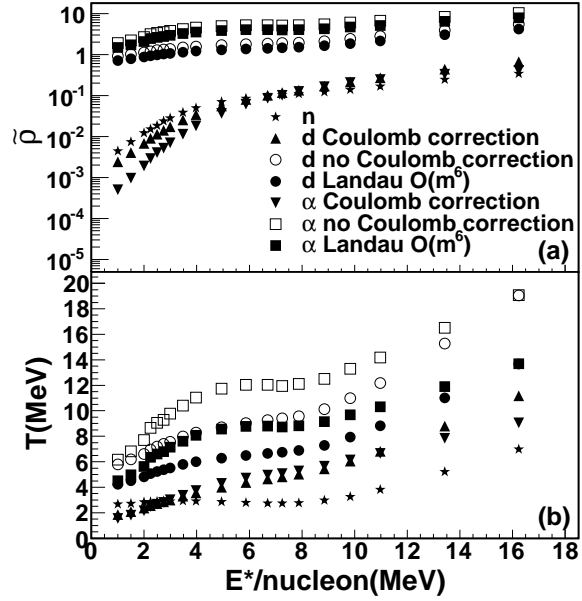}
\caption{The reduced density (a) and temperature $T$ (b) versus E*/nucleon of $d$ and $\alpha$ from CoMD simulations. Three methods, with Coulomb correction, without Coulomb correction and Landau's $O(m^6)$ theory, are used to calculate the density and temperature. The  corresponding results for neutrons are also included as a reference.}
\label{try4}
\end{figure}

It is instructive to study the behavior of the quadrupole fluctuations and the density below the critical point.  In figure 3 we plot these quantities divided by their values at the critical temperature as function of $T/T_c$. The behavior when the Coulomb term is zero $(A'=0)$ was already discussed in \cite{hua3}. For finite Coulomb potential, we observe that the number of condensate bosons increases faster with decreasing $T$ and, accordingly, the quadrupole fluctuations are larger. In fact the larger the Coulomb repulsion, the higher are the fluctuations already at zero $T$, which is intuitively clear: particles emitted from a source at zero $T$, will develop substantial final momenta because of the large Coulomb acceleration and consequently will exhibit large fluctuations.  Naturally, we have to keep in mind that at zero $T$, bosons might be confined by an attractive mean field. 

In figure \ref{try4} we plot $\tilde\rho=\frac{\rho}{\rho_0}$ where $\rho_0$ is the nuclear ground state density and $T$ vs excitation energy respectively obtained from CoMD simulations. The neutron case is also included \cite{hua1, hua2}. As we see the densities derived from  $d$ and $\alpha$ observables  with Coulomb correction are very close to each other and to those derived from the the neutron observables.There is a large difference between the cases with Coulomb correction and without Coulomb correction  which demonstrates the crucial role of adding the Coulomb repulsion between bosons. For completeness we also include the results for bosons from Landau's $O(m^6)$ approach \cite{hua3} which is close to the results without Coulomb corrections. The derived $T$ of $d$ and $\alpha$ with Coulomb correction are also much closer to the neutron values. The good agreement for the obtained temperatures and densities suggests that thermal equilibrium is nearly reached for particles emitted in the transverse direction 
\begin{figure}
\centering
\includegraphics[width=0.5\columnwidth]{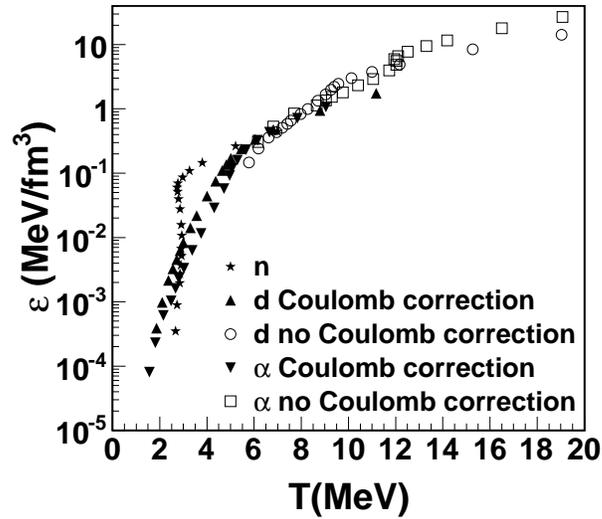} 
\caption{Energy density versus T for different cases, see figure \ref{try4}.}
\label{try5}
\end{figure}

\begin{figure}
\centering
\includegraphics[width=0.5\columnwidth]{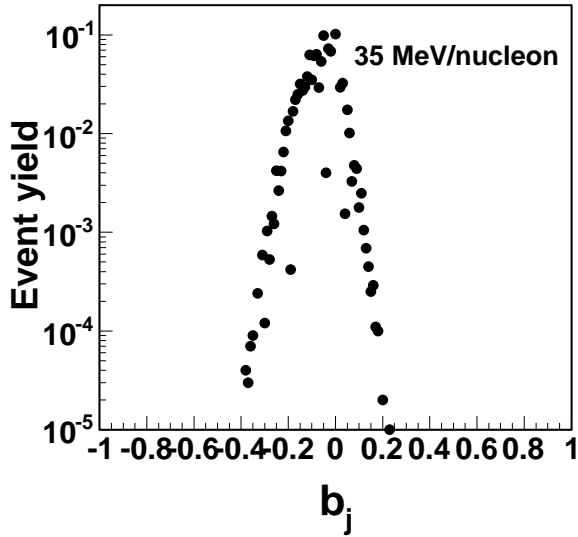} 
\caption{The $b_j$ distribution for CoMD at $35$ MeV/nucleon.}
\label{try6}
\end{figure}

In figure \ref{try5}, we plot the energy density $\varepsilon=\frac{E}{N}\rho$ versus $T$.  Without Coulomb corrections, the results are systematically located at larger $T$ and energy density with respect to $n$. When the correction is included, we obtain a curve very close to that derived from the neutron data which   do not reflect Coulomb effects (at least not directly).  Furthermore they are fermions. This result shows that when all the different effects are properly taken into account, we obtain a unique energy density behavior which demonstrates that different particles experience a sudden increase of the degrees of freedom (fragmentation) at the same $T$ about 4 MeV. At higher $T$ the energy density increases because of a liquid-gas phase transition. It is important to notice that our results but also other results in the literature \cite{justin, kris, elliott, pochodzalla, borderie} seem to give always $T$ smaller than the nuclear matter temperature at the critical point for a liquid-gas phase transition. The impossibility of reaching $T_{cLG}$ was also predicted in microscopic dynamical calculations of the Lyapunov exponents. In those calculations, the Lyapunov exponents will not increase beyond a certain value since collective effects will set in \cite{aldo3}.  We stress that the critical temperature for a liquid-gas phase transition has nothing to do, in principle, with the critical point of a Bose condensate. The CoMD model used in this work has no knowledge of a Bose condensate, thus we do not expect any spectacular effects to be observable in figure 5. However, there are many signatures of a possible condensate in nuclei even though none has been so far conclusive. We mention the Hoyle state in ${}^{12}C$, but also the anomalous large number of $\alpha$ particles in the universe as compared to hydrogen \cite{bbn1, bbn2}. In heavy ion collisions a large production of $\alpha$ is observed. Recently, some experimental signatures of a condensate have been proposed \cite{raduta, mandredi}. In the calculations discussed here we have always implicitly assumed that the number of bosons is constant, which is crucial to have a condensate. In reality, during the collisions, even though we might start from 'perfect' $\alpha$ cluster nuclei, because of the large excitation energy, $\alpha$ particles might be destroyed and thus we obtain in general a mixture of fermions and bosons. This is of course especially severe for $d$-like events.  In order to avoid this problem we propose the following strategy to select 'good' events. First we define the quantity:
\begin{equation}
b_j= \frac{1}{M}\sum_{i=1}^M \frac{(-1)^{Z_i}+(-1)^{N_i}}{2},
\end{equation}
where $M$ is the multiplicity in one event, $Z_i$ and $N_i$ are the proton and neutron number in the $i^{th}$ fragment in that event respectively. The meaning of such a quantity is clear: if the final fragments for instance are all $d$-like, we get $b_j=-1$, while for pure $\alpha$ like fragments $b_j=+1$, if we further select N=Z nuclei. Pure fermion cases give $b_j=0$. In figure \ref{try6} we plot the $b_j$ distribution from CoMD calculations. As we see in the figure the model gives an average $b_j$ close to zero, which means that most of the final fragments in the model are 'fermion-like'. Recall that the model takes into account mainly the Pauli principle. However, preliminary experimental results on ${}^{40}Ca+{}^{40}Ca$ collisions \cite{kasia, indra}, display much larger distributions than in figure \ref{try6}. In particular events are observed near $b_j=\pm 1$ which could be a signature for a Bose condensate. Therefore, we propose to select fragments from data with $b_j=1 (-1)$, $N=Z$, and perform the analysis discussed in this paper to obtain the density and temperature of the bosons for each excitation energy. The energy density might be constructed for different situations and compared to fermions. 

\section{Conclusion}
In conclusion, in this paper we have discussed Coulomb modifications to the density and temperature in heavy ion collisions. The quantum case (bosons) has been discussed. We have shown that the temperatures obtained from different particle types are very similar to those derived from neutron observables, which implies the 'near ergodicity' of the system. The energy densities are very similar at high temperatures, which suggests that Coulomb corrections are small due to the small source densities. Experimental investigations of the effects discussed in this work for well determined sources and excitation energies \cite{alan1, sara, justin, qin, borderie, kasia} would be very important to further constrain the Nuclear Equation of State in the liquid-gas phase transition region also for asymmetric matter. We suggest a selection of  data according to excitation energy and to their $b_j$ distribution as defined in this paper. The $T$, $\rho$ and energy density for $d$-like, $\alpha$-like and fermion like events might be compared to pin down the possibility of a condensate in nuclei. \\

We thank prof. J. Natowitz for discussions and a critical reading of the manuscript. A mathematica code for the quantum case is available from the authors upon request.

\appendix
\section{} \label{quanboson}
For the quantum case, assuming particles follow the Bose-Einstein distribution modified by the Coulomb correction,
\begin{equation}
f(p)=\frac{1}{e^{[\varepsilon+\frac{1.44\times4\pi\hbar^2 Z_dZ_s}{Vp^2}-\mu]/T}-1},
\end{equation}
where $\varepsilon=\frac{p^2}{2m}$ is the energy , $\mu$ is the chemical potential, $T$ is the temperature.
The average number of particles is
\begin{eqnarray}
N &=&\frac{g}{h^3}\int d^3xd^3p f(p)\nonumber\\
&=&\frac{gV}{h^3}4\pi\int_0^\infty dpp^2 f(p) .
\label{av_particle1}
\end{eqnarray}
where $g$ is the degeneracy. Let's make the integral variable transformation,
\begin{equation}
\varepsilon=\frac{p^2}{2m},\quad p=(2m\varepsilon)^{\frac{1}{2}}, \quad dp=\frac{m}{\sqrt{2m\varepsilon}}d\varepsilon. \label{transformation1}
\end{equation}
Thus Eq. (\ref{av_particle1}) becomes
\begin{eqnarray}
N &=&\frac{gV}{h^3}4\pi\int_0^\infty dpp^2f(p)\nonumber\\
&=&\frac{gV}{h^3}4\pi\frac{(2m)^\frac{3}{2}}{2}\int_0^\infty d\varepsilon \varepsilon^\frac{1}{2} f(\varepsilon)\nonumber\\
&=&\frac{gV}{h^3}4\pi\frac{(2m)^\frac{3}{2}}{2}\int_0^\infty d\varepsilon \varepsilon^\frac{1}{2}\frac{1}{e^{[\varepsilon+ \frac{1.44\times4\pi\hbar^2 Z_dZ_s}{Vp^2}-\mu]/T}-1}\nonumber\\
&=&\frac{gV}{h^3}4\pi\frac{(2m)^\frac{3}{2}}{2}\int_0^\infty d\varepsilon \varepsilon^\frac{1}{2}\frac{1}{e^{[\varepsilon+ \frac{1.44\times4\pi\hbar^2 Z_dZ_s}{2mV\varepsilon}-\mu]/T}-1}\nonumber\\
&=&\frac{gV}{h^3}4\pi\frac{(2m)^\frac{3}{2}}{2}\int_0^\infty d\varepsilon \varepsilon^\frac{1}{2}\frac{1}{e^{[\varepsilon+ \frac{A'}{V\varepsilon}-\mu]/T}-1},
\label{av_particle2}
\end{eqnarray}
where $A'=\frac{1.44\times4\pi\hbar^2 Z_dZ_s}{2m}$. Define:
\begin{equation}
y=\frac{\varepsilon}{T}, \quad \nu=\frac{\mu}{T}.\label{notation}
\end{equation}
Therefore, Eq. (\ref{av_particle2}) becomes
\begin{equation}
N =\frac{gV}{h^3}4\pi\frac{(2mT)^\frac{3}{2}}{2}\int_0^\infty dy y^\frac{1}{2}\frac{1}{e^{y+\frac{A'}{yVT^2}-\nu}-1}.
\label{av_particle3}
\end{equation}
When $T>T_c$, the multiplicity fluctuation is
\begin{equation}
\langle(\Delta N)^2\rangle=T(\frac{\partial N}{\partial \mu})_{T, V}=(\frac{\partial N}{\partial \nu})_{T, V}.\label{fluctuation1}
\end{equation}
Substitute Eq. (\ref{av_particle3}) into Eq. (\ref{fluctuation1}), we get:
\begin{eqnarray}
\langle(\Delta N)^2\rangle=\frac{gV}{h^3}4\pi\frac{(2mT)^\frac{3}{2}}{2}\int_0^\infty dy y^\frac{1}{2}\frac{e^{y+\frac{A'}{yVT^2}-\nu}}{(e^{y+\frac{A'}{yVT^2}-\nu}-1)^2}.\label{fluctuation3}
\end{eqnarray}
Dividing Eq. (\ref{fluctuation3}) by Eq. (\ref{av_particle3}) gives
\begin{eqnarray}
\frac{\langle(\Delta N)^2\rangle}{N}&=&\frac{\frac{gV}{h^3}4\pi\frac{(2mT)^\frac{3}{2}}{2}\int_0^\infty dy y^\frac{1}{2}\frac{e^{y+\frac{A'}{yVT^2}-\nu}}{(e^{y+\frac{A}{yT^2}-\nu}-1)^2}}{\frac{gV}{h^3}4\pi\frac{(2mT)^\frac{3}{2}}{2}\int_0^\infty dy y^\frac{1}{2}\frac{1}{e^{y+\frac{A'}{yVT^2}-\nu}-1}}\nonumber\\
&=&\frac{\int_0^\infty dy y^\frac{1}{2}\frac{e^{y+\frac{A}{yT^2}-\nu}}{(e^{y+\frac{A}{yT^2}-\nu}-1)^2}}{\int_0^\infty dy y^\frac{1}{2}\frac{1}{e^{y+\frac{A}{yT^2}-\nu}-1}}.
\label{fluctuation4}
\end{eqnarray}
In the same framework, we also calculate the quadrupole momentum fluctuation: 
\begin{eqnarray}
\langle \sigma_{xy}^2 \rangle&=&\frac{\int d^3p (p_x^2-p_y^2)^2 \frac{1}{e^{[\frac{p^2}{2m}+\frac{1.44\times 4\pi\hbar^2 Z_pZ_s }{p^2V}-\mu]/T}-1}}{\int d^3p \frac{1}{e^{[\frac{p^2}{2m}+\frac{1.44\times 4\pi\hbar^2 Z_pZ_s }{p^2V}-\mu]/T}-1}}\nonumber\\
&=&(2mT)^2\frac{4}{15}\frac{\int_0^\infty dy y^\frac{5}{2} \frac{1}{e^{y+\frac{A'}{yVT^2}-\nu}-1}}{\int_0^\infty dy y^\frac{1}{2} \frac{1}{e^{y+\frac{A'}{yVT^2}-\nu}-1}}
\label{quadrupole4}
\end{eqnarray}



\end{document}